\documentclass[journal]{IEEETran}

\IEEEoverridecommandlockouts
\usepackage{indentfirst,flushend}
\usepackage{amssymb,bm,mathrsfs,times,amscd,amsmath,amsthm,amsfonts,bbm}
\usepackage{latexsym}
\usepackage{dsfont}
\usepackage{graphicx}
\usepackage{subfigure}
\usepackage{color}
\usepackage{cases}

\newtheorem{remark}{Remark}
\usepackage{multirow}
\usepackage[sort]{cite}
\usepackage{multicol}
\usepackage{clrscode}
\usepackage{algorithm}
\usepackage{psfrag,stfloats,nicefrac}
\usepackage{algorithmic}
\usepackage{supertabular}
\usepackage{makecell}
\usepackage{booktabs}
\usepackage{threeparttable}
\usepackage[justification=centering]{caption}
\usepackage{epstopdf}
\usepackage{subfigure}
\allowdisplaybreaks[4]

\begin{document}
\title{Maximizing the Set Cardinality of Users Scheduled for Ultra-dense uRLLC Networks}
\author{Shiwen~He,~\IEEEmembership{Member,~IEEE},~Jun Yuan,~Zhenyu An,~\IEEEmembership{Student Member,~IEEE},~Yunshan Yi,~Yongming Huang,~\IEEEmembership{Senior Member,~IEEE}
\thanks{S. He and J. Yuan are with the School of Computer Science and Engineering, Central South University, Changsha 410083, China. S. He is also with the Purple Mountain Laboratories, Nanjing 210096, China. (email: \{shiwen.he.hn, yuanjun\}@csu.edu.cn). }
\thanks{Z. An and Y. Yi are with the Purple Mountain Laboratories, Nanjing 210096, China. (email: \{anzhenyu, yiyunshan\}@pmlabs.com.cn). }
\thanks{Y. Huang is with the National Mobile Communications Research Laboratory, School of Information Science and Engineering, Southeast University, Nanjing 210096, China. He is also with the Purple Mountain Laboratories, Nanjing 210096, China. (email: huangym@seu.edu.cn). }
}

\maketitle
\vspace{-.6 in}

\begin{abstract}
Ultra-reliability and low latency communication has long been an important but challenging task in the fifth and sixth generation wireless communication systems. Scheduling as many users as possible to serve on the limited time-frequency resource is one of a crucial topic, subjecting to the maximum allowable transmission power and the minimum rate requirement of each user. We address it by proposing a mixed integer programming model, with the goal of  maximizing the set cardinality of users instead of maximizing the system sum rate or energy efficiency. Mathematical transformations and successive convex approximation are combined to solve the complex optimization problem. Numerical results show that the proposed method achieves a considerable performance compared with exhaustive search method, but with lower computational complexity.
\end{abstract}

\section{Introduction}
Ultra-reliability and low latency communication (uRLLC) requires $99.999\%$ reliability and $1$ ms latency~\cite{3GPP} for transmitting a $32$ bytes packet. Recently, discussion on the property of achievable rate with finite blocklength transmission is one hot topic for uRLLC~\cite{TITPoly2010}.  The authors of~\cite{Ren2020} studied low bound of blocklength and applied it to power allocation with four different downlink transmission schemes. The authors of~\cite{Pan2019} proposed a novel perturbation-based iterative algorithm to jointly optimize the blocklength allocation and the unmanned aerial vehicle's location. The authors of~\cite{TWCHe2020} obtained an analytical condition under which the minimum rate requirement of any user can be satisfied. The authors of~\cite{TWCNasir2021} consider three different problems with the objective of maximizing the users' minimum rate.

For multiuser multi-antenna  communication systems,  a popular topic is joint beamforming (BF) and user scheduling (US), which could be formulated as non-convex mixed integer programming problem. It was usually studied based on conventional Shannon capacity~\cite{BellShannon1948}. The authors of~\cite{Chen2017} proposed a low complexity BF and US scheme for the downlink multiple-input multiple-output based non-orthogonal multiple access (MIMO-NOMA) system, but the non-convex problem was handled by independently tracking two subproblems, namely, BF scheme and greedy min-power US scheme, instead of jointly solving them. The authors of~\cite{Jiang2018} investigated the joint US and analog beam selection problem for codebook-based massive MIMO downlink systems with hybrid antenna architecture. Its optimization objective function was to maximizing the user rate with allowable transmission power constraint and a $0-1$ diagonal digital precoding matrix, but without considering the demand of minimum rate of each user. The authors of~\cite{Chen2021} considered the joint coordinated BF and US for coordinated multi-point enabled new radio in unlicensed spectrum networks, with also aiming at maximizing the throughput instead of considering the rate demand for each user. {An alternative approach for user scheduling is formulated as a series of decision-making problems, where the user relationship between the previous and current time is considered. More recently, the domain knowledge and deep learning~\cite{She2021, Gu2021} are integrated and adopted to solve the problem of user scheduling, but few studies investigated how to maximize the set cardinality of scheduled users.}

For ultra-dense uRLLC systems, compared to maximizing the system sum rate, scheduling as many users as possible to be served on the same time-frequency resource is more important under the maximum allowable transmission power and the minimum rate requirement of each user.
Motivated by these observations, in the letter, we aim at scheduling as many serving users as possible via optimizing the transmitting precoding simultaneously subject to the maximum allowable transmission power and the minimum rate requirement of each user.
A resource allocation model which simultaneously considering US and BF is proposed. Mathematical transformations and successive convex approximation are combined to solve the original integer mixed optimization problem. Finally, simulation results reveal its effectiveness.

\section{\label{introduction}System Model and Problem Formulation}


Consider a downlink multiple-input single-output uRLLC system with one BS and $K$ single antenna users. BS is equipped with $N_{\mathrm{t}}$ antenna and $K \ge N_{\mathrm{t}}$. Let $\mathcal{K}=\left\{1,2,\cdots,K\right\}$ and $\mathcal{S}\subseteq\mathcal{K}$ be the set of all users and served users, respectively. Let $\mathbf{w}_{k}\in\mathbb{C}^{N_{\mathrm{t}}\times 1}$, $s_{k}$ and $\mathbf{h}_{k}\in\mathbb{C}^{N_{\mathrm{t}}\times 1}$ respectively denote the beamforming vector used by the BS, the baseband signal for the $k$-th user, and the channel coefficient between the BS and the $k$-th user.

The received baseband signal at the $k$-th user is expressed as:
\begin{equation}\label{URLLC01}
y_{k}=\sum\limits_{l\in\mathcal{S}}\mathbf{h}_{k}^{H}\mathbf{w}_{l}s_{l}+n_{k}{,}
\end{equation}
where $n_{k}$ is the zero-mean additive white Gaussian noise with variance $\sigma_{k}^{2}$ at the $k$-th user. Thus, the signal-to-interference-plus-noise ratio (SINR) of the $k$-th user is calculated as~\eqref{URLLC02} with $\overline{\mathbf{h}}_{k}=\frac{\mathbf{h}_{k}}{\sigma_{k}}$.
\begin{equation}\label{URLLC02}
\gamma_{k}=\frac{\left|\overline{\mathbf{h}}_{k}^{H}\mathbf{w}_{k}\right|^{2}}
{\sum\limits_{l\neq k,l\in\mathcal{S}}\left|\overline{\mathbf{h}}_{k}^{H}\mathbf{w}_{l}\right|^{2}+1}.
\end{equation}
{The achievable rate obtained by taking the parameters of decoding error probability $\epsilon$ and finite blocklength $n$ ( or $n$ channels used for transmitting the user data) under consideration is described as~\cite{TITPoly2010,TWCNasir2021}}
\begin{equation}\label{URLLC03}
R\left(\gamma_{k}\right)=C\left(\gamma_{k}\right)-\vartheta\sqrt{V\left(\gamma_{k}\right)},
\end{equation}
where $C\left(\gamma\right)=\ln\left(1+\gamma\right)$,  $\vartheta=\frac{Q^{-1}\left(\epsilon\right)}{\sqrt{n}}$ and $Q^{-1}\left(\cdot\right)$ is the inverse of Gaussian Q-function $Q\left(x\right)=\frac{1}{\sqrt{2\pi}}\int_{x}^{\infty}\mathrm{exp}\left(-\frac{t^{2}}{2}\right)dt$, and $V\left(\gamma\right)$ is defined as
\begin{equation}\label{URLLC04}
V\left(\gamma\right)=1-\frac{1}{\left(1+\gamma\right)^{2}}.
\end{equation}
It is not difficult to find that compared to the classical Shannon capacity\footnote{The proposed algorithm in the letter is also suitable for communication with Shannon capacity.}, this is a more conservative description of channel capacity. For the simplicity, we assume that the $n$ channels used remain unchanged, that is, the channel model used is a slow time-varying channel.

{The set $\mathcal{S}$ of served users should be carefully scheduled in order to effectively achieve the multiuser diversity gain and assure each user obtain a minimum achievable rate i.e., $R\left(\gamma_{k}\right)\geq r_{k} \geq \frac{D}{n}$, where $D$ represents transmitting data size. Note that $D$ bits data is equally divided to $n$ channels since the slow time-varying channel is assumed in the letter.} To schedule as many users as possible under the condition of the requirement of minimum user rate and the maximum allowable transmitting power, the joint BF and US problem is formulated as follows
\begin{subequations}\label{URLLC05}
\begin{align}
&\max_{\mathcal{S}\subseteq\mathcal{K}, \left\{\mathbf{w}_{k}\right\}} \left|\mathcal{S}\right|, \label{URLLC05a}\\
\mathrm{s.t.}&~r_{k}\leq R\left(\gamma_{k}\right), \forall k\in\mathcal{S},\label{URLLC05b}\\
&\sum\limits_{k\in\mathcal{S}}\left\|\mathbf{w}_{k}\right\|_{2}^{2}\leq P.\label{URLLC05c}
\end{align}
\end{subequations}
In~\eqref{URLLC05a}, $\left|\mathcal{S}\right|$ denotes the cardinality of set $\mathcal{S}$ { and $P$ is the maximum allowable transmission power of base station}.  Note that problem~\eqref{URLLC05} is non-convex problem and thus is hard to obtain the solution directly. Furthermore, due to the uncertainty of user scheduling set $\mathcal{S}$, problem~\eqref{URLLC05} is essentially an uncertainty problem. Generally speaking, brute exhaustive searching may be an effective selection for solving the considered problem at the cost of high computational complexity, especially for a large {number} of users to be scheduled.

\section{\label{AlgorithDesign} Design of Optimization Algorithm}

To release the uncertainty of user scheduling set $\mathcal{S}$ and obtain a tractable form of problem~\eqref{URLLC05}, we introduce auxiliary variable $\kappa_{k}$ to be an indicator of user scheduling. Specifically, if $\kappa_{k}=1$, the $k$-th user is scheduled as a communication user, i.e., $k\in\mathcal{S}$. Otherwise, $\kappa_{k}=0$ means that the $k$-th user is not scheduled as a communication user, i.e.,  $k\notin\mathcal{S}$. Let $\mathbf{\kappa}=\left[\kappa_{1},\kappa_{2},\cdots,\kappa_{k},\cdots,\kappa_{K}\right]^{T}$, thus, we can rewrite equivalently problem~\eqref{URLLC05} as follows
\begin{subequations}\label{URLLC06}
\begin{align}
&\max_{\left\{\kappa_{k},\mathbf{w}_{k}\right\}} \left\|\mathbf{\kappa}\right\|_{0}, \label{URLLC06a}\\
\mathrm{s.t.}&~\kappa_{k}r_{k}\leq R\left(\gamma_{k}\right), \forall k\in\mathcal{K},\label{URLLC06b}\\
&\kappa_{k}\in\{0,1\}, \forall k\in\mathcal{K},\label{URLLC06c}\\
&\sum\limits_{k\in\mathcal{K}}\left\|\mathbf{w}_{k}\right\|_{2}^{2}\leq P. \label{URLLC06d}
\end{align}
\end{subequations}
Problem~\eqref{URLLC06} is an equivalent form of problem~\eqref{URLLC05} and the main goal of objective functions~\eqref{URLLC06a} is to assure that as many users as possible are scheduled. Obviously, achievable rate $R\left(\gamma_{k}\right)$ in constraint ~\eqref{URLLC06b} is non-convex nor non-concave, with a square root and inverse form of $\gamma _k$.  Besides, constraint~\eqref{URLLC06c} make it be a mixed integer continuous programming problem and also brings difficulties.  Therefore, it is very difficult to obtain the global optimal solution of problem~(6), even the local optimal solution.

In what follows, we focus on designing an effective iterative optimization algorithm to solve problem~\eqref{URLLC06} using a series of basic mathematical operations. {According to Theorem 2 and Corollary 1  in reference~\cite{TWCHe2020}, the required minimum SINR for satisfying the requirement of minimum user rate is easily obtained with an analytical solution. Consequently, constraint~\eqref{URLLC06b} could be equivalently transformed to the following form:}
\begin{subequations}\label{URLLC07}
\begin{align}
&\max_{\left\{\kappa_{k},\mathbf{w}_{k}\right\}} \left\|\mathbf{\kappa}\right\|_{0}, \label{URLLC07a}\\
\mathrm{s.t.}&~\kappa_{k}\widetilde{\gamma}_{k}\leq \gamma_{k}, \forall k\in\mathcal{K},~\eqref{URLLC06c},~\eqref{URLLC06d},\label{URLLC07b}
\end{align}
\end{subequations}
where $\widetilde{\gamma}_{k}$ is the minimum SINR for achieving the minimum rate $r_{k}$ and can be obtained via using the conclusion of~\cite[Corollary 1]{TWCHe2020}. Problem~\eqref{URLLC07} can be transformed into the following form by simultaneously substituting~\eqref{URLLC02} to~\eqref{URLLC07b} and introducing an auxiliary variable $\varphi_k$.
\begin{subequations}\label{URLLC08}
\begin{align}
&\max_{\left\{\kappa_{k},\varphi_{k},\mathbf{w}_{k}\right\}} \sum\limits_{k\in\mathcal{K}}\kappa_{k}, \label{URLLC08a}\\
\mathrm{s.t.}~&0\leq\kappa_{k}\leq 1, \forall k\in\mathcal{K},\label{URLLC08b}\\
&\sum\limits_{k\in\mathcal{K}}\left(\kappa_{k}-\kappa_{k}^{2}\right)\leq 0,\label{URLLC08c}\\
&\kappa_{k}\widetilde{\gamma}_{k}- \frac{\left|\overline{\mathbf{h}}_{k}^{H}\mathbf{w}_{k}\right|^{2}}{\varphi_{k}}\leq 0, \forall k\in\mathcal{K},\label{URLLC08d}\\
&\sum\limits_{l\neq k,l\in\mathcal{K}}\left|\overline{\mathbf{h}}_{k}^{H}\mathbf{w}_{l}\right|^{2}+1\leq \varphi_{k}, \forall k\in\mathcal{K},~\eqref{URLLC06d}, \label{URLLC08e}
\end{align}
\end{subequations}
{where constraints~\eqref{URLLC08b} and~\eqref{URLLC08c} are equivalent to~\eqref{URLLC06c}. Actually,~\eqref{URLLC08b} limits $\kappa_{k}$ value between 0 and 1, and the inequality~\eqref{URLLC08b} holds if and only if every $\kappa _k$ is 0 or 1 (any $\kappa _k$ with value between 0 and 1 will make the summation be larger than 0). Note that the summation in denominator of (2) is expanded from set $\mathcal{S}$ to $\mathcal{K}$.
According to \cite[Proposition 2]{Che2014}, the strong Lagrangian duality holds for the above problem \eqref{URLLC08}. Using similar mathematical tricks on handling ${(\sum\limits_{k \in {\cal K}} {{\kappa _k}} )^2}$, we resort to solve problem~\eqref{URLLC08} via solving problem~\eqref{URLLC09} by appropriately choosing $\mu>0$.}
\begin{subequations}\label{URLLC09}
\begin{align}
&\min_{\left\{\kappa_{k},\varphi_{k},\mathbf{w}_{k}\right\}}-\sum\limits_{k\in\mathcal{K}}\kappa_{k}+g\left(\mathbf{\kappa}\right)-h\left(\mathbf{\kappa}\right), \label{URLLC09a}\\
\mathrm{s.t.}~&\eqref{URLLC08b},~\eqref{URLLC08d},~\eqref{URLLC08e},~\eqref{URLLC06d},\label{URLLC09b}
\end{align}
\end{subequations}
where $g\left(\mathbf{\kappa}\right)$ and $h\left(\mathbf{\kappa}\right)$ are defined respectively as
\begin{subequations}\label{URLLC10}
\begin{align}
g\left(\mathbf{\kappa}\right)&\triangleq\mu\sum\limits_{k\in\mathcal{K}}\kappa_{k}+\mu\left(\sum\limits_{k\in\mathcal{K}}\kappa_{k}\right)^{2}, \label{URLLC10a}\\
h\left(\mathbf{\kappa}\right)&\triangleq\mu\sum\limits_{k\in\mathcal{K}}\kappa_{k}^{2}+\mu\left(\sum\limits_{k\in\mathcal{K}}\kappa_{k}\right)^{2}.\label{URLLC10b}
\end{align}
\end{subequations}
For fixed $\mu$, {~\eqref{URLLC09a} and~\eqref{URLLC08d} are non-convex since the differences of convex form, which brings difficulties for solving problem~\eqref{URLLC09}.} Fortunately, note that function $\kappa_{k}^{2}$ is convex, while function $\frac{\left|\mathbf{h}_{k}^{H}\mathbf{w}_{k}\right|^{2}}{\varphi_{k}}$ is quadratic-over-affine function, which is jointly convex w.r.t. the involved variables.

In the sequel, we resort to solve problem~\eqref{URLLC09} via using successive convex approximation methods~\cite{He2017}. Using the convexity of function $h\left(\mathbf{\kappa}\right)$, we have
\begin{equation}\label{URLLC11}
h\left(\mathbf{\kappa}\right)\geq h\left(\kappa^{\left(\tau\right)}\right)+\sum\limits_{k\in\mathcal{K}}h'\left(\kappa_{k}^{\left(\tau\right)}\right)\left(\kappa_{k}-\kappa_{k}^{\left(\tau\right)}\right)\triangleq\phi\left(\mathbf{\kappa}\right)
\end{equation}
where $h'\left(\kappa_{k}\right)=2\mu\left(\kappa_{k}+\left(\sum\limits_{k\in\mathcal{K}}\kappa_{k}\right)\right)$ and the superscript $\tau$ denotes the $\tau$-th iteration of the iterative algorithm presented shortly. Note that $\phi\left(\mathbf{\kappa}\right)$ is in fact the first-order Taylor series expansion approximation of function $h\left(\mathbf{\kappa}\right)$ around the point $\kappa_{k}^{\left(\tau\right)}$. Similarly, using the convex property of function $\frac{\left|\mathbf{h}_{k}^{H}\mathbf{w}_{k}\right|^{2}}{\varphi_{k}}$, we can obtain their low boundary approximation as follows
\begin{equation}\label{URLLC12}
\begin{split}
&\frac{\left|\mathbf{h}_{k}^{H}\mathbf{w}_{k}\right|^{2}}{\varphi_{k}}\geq
\overline{\varphi}\left(\mathbf{w}_{k},\varphi_{k}\right)\\
&\triangleq\frac{2\Re\left(\left(
\mathbf{w}_{k}^{\left(\tau\right)}\right)^{H}\mathbf{h}_{k}\mathbf{h}_{k}^{H}\mathbf{w}_{k}\right)}
{\varphi_{k}^{\left(\tau\right)}}-\left(\frac{\left|\mathbf{h}_{k}^{H}\mathbf{w}_{k}^{\left(\tau\right)}\right|}
{\varphi_{k}^{\left(\tau\right)}}\right)^{2}\varphi_{k}.
\end{split}
\end{equation}
From the aforementioned discussions, the convex approximation problem solved at the $\left(\tau+1\right)$-th iteration of the proposed algorithm is given by
\begin{subequations}\label{URLLC13}
\begin{align}
&\min_{\left\{\kappa_{k},\varphi_{k},\mathbf{w}_{k}\right\}} -\sum\limits_{k\in\mathcal{K}}\kappa_{k}+g\left(\mathbf{\kappa}\right)-\phi\left(\mathbf{\kappa}\right), \label{URLLC12a}\\
\mathrm{s.t.}~&\eqref{URLLC08b}, \eqref{URLLC08e}, \eqref{URLLC06d},\label{URLLC12b}\\
&\kappa_{k}\widetilde{\gamma}_{k}- \overline{\varphi}\left(\mathbf{w}_{k},\varphi_{k}\right)\leq 0, \forall k\in\mathcal{K}.\label{URLLC12c}
\end{align}
\end{subequations}
Note that the objective function of problem~\eqref{URLLC13} is an upper boundary of the objective function of problem~\eqref{URLLC09}, while the feasible set of problem~\eqref{URLLC13} is a subset of the feasible set of problem~\eqref{URLLC09}. Consequently, the solution  obtained via problem~\eqref{URLLC13} is an upper boundary of problem~\eqref{URLLC09}. {Here, CVX \cite{CVX2020} is adopted to finally obtain the solution.}
\begin{algorithm}[htp]
\caption{Solution of problem~\eqref{URLLC13}}\label{URLLCA01}
\begin{algorithmic}[1]
\STATE Let $\tau=0$. Initialize beamforming vector $\mathbf{w}_{k}^{\left(0\right)}$, $\forall k\in\mathcal{K}$, {$\mu = 0.05$}, $\delta = 10^{-3}$, and $\kappa_{k}^{\left(0\right)}$, $\forall k\in\mathcal{K}$, such that constraint~\eqref{URLLC07b} is satisfied.\label{URLLCA0101}
\STATE Initialize $\varsigma^{\left(0\right)}$ and $\varphi_{k}^{\left(0\right)}$.\label{URLLCA0102}
\STATE Let $\tau\leftarrow \tau+1$. Solve problem~\eqref{URLLC13} to obtain $\kappa_{k}^{\left(\tau\right)}$, $\varphi_{k}^{\left(\tau\right)}$, and $\mathbf{w}_{k}^{\left(\tau\right)}$, $\forall k\in\mathcal{K}$. \label{URLLCA0103}
\STATE Calculate objective value $\varsigma^{(\tau)}$. If $\left|\frac{\varsigma^{\left(\tau\right)}-\varsigma^{\left(\tau-1\right)}}{\varsigma^{\left(\tau-1\right)}}\right|\leq\delta$, stop iteration. Otherwise, go to step~\ref{URLLCA0103}.  \label{URLLCA0104}
\end{algorithmic}
\end{algorithm}

An iterative algorithm summarized as Algorithm 1 is outlined for solving problem~\eqref{URLLC13}, where $\varsigma^{\left(\tau\right)}$  denotes the objective value of optimization problem~\eqref{URLLC13} at the $\tau$-th iteration iteration. To speed up the convergence of Algorithm~\ref{URLLCA01}, we can first filter out the users who meet constraints~\eqref{URLLC05b} and~\eqref{URLLC05c} by {solving single user communication with maximum ratio transmission and full power transmission.}
\begin{figure*} [ht]\centering
\subfigure[n = 128, $\epsilon$ = 1e-6.] {
 \label{fig:a}
\includegraphics[width=0.62\columnwidth]{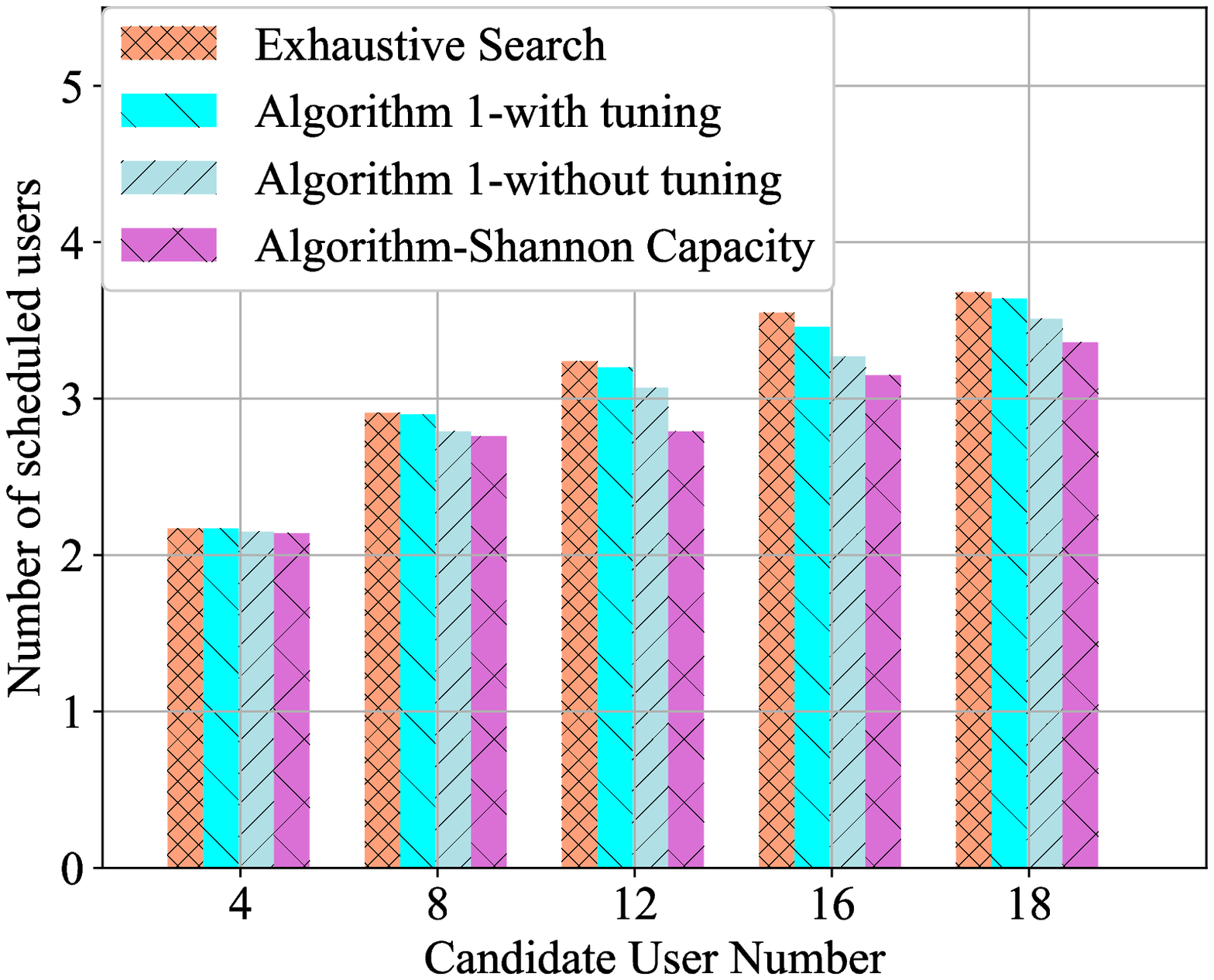}
}
\subfigure[K = 8, $\epsilon$ = 1e-6.] {
\label{fig:b}
\includegraphics[width=0.62\columnwidth]{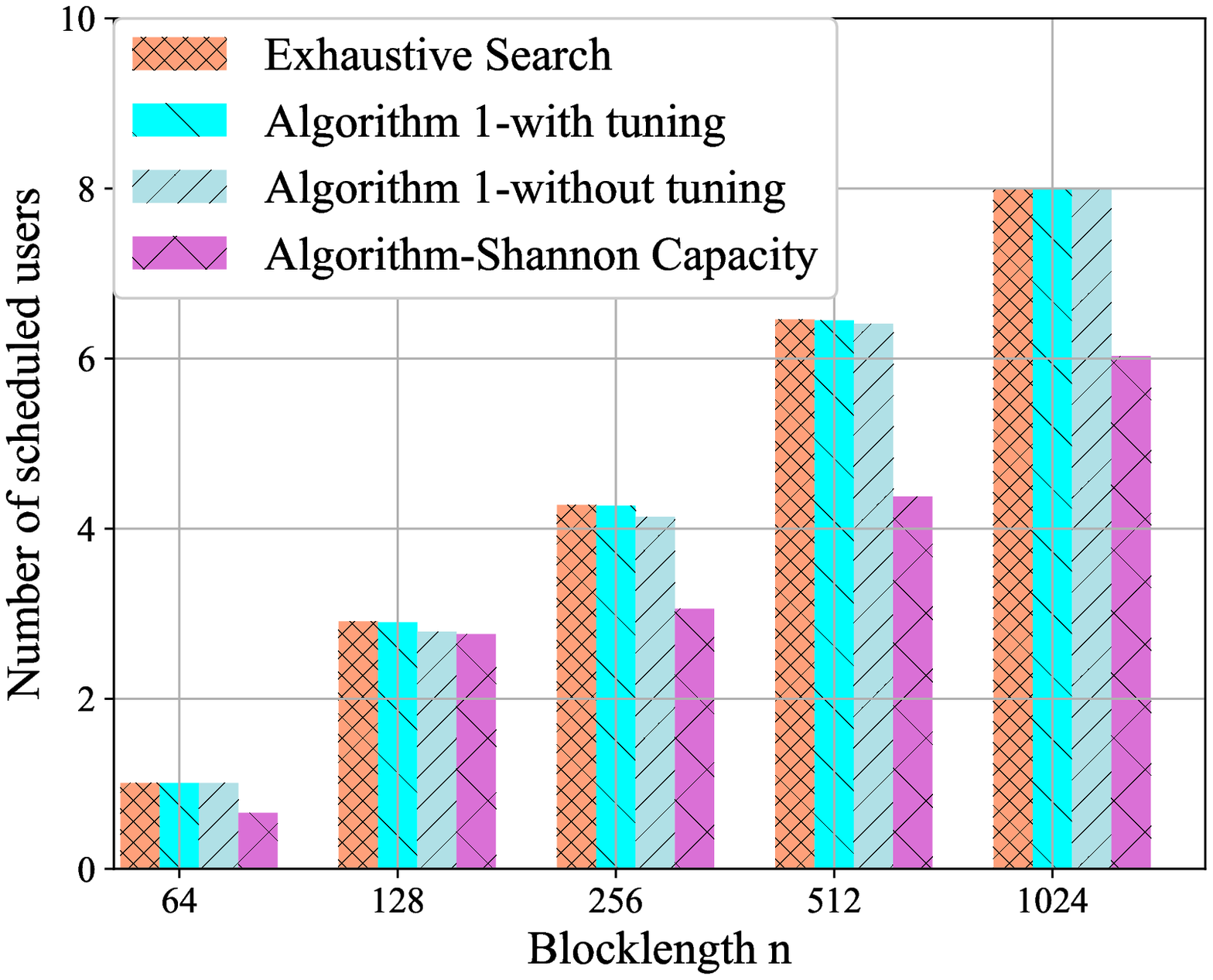}
}
\subfigure[K = 8, n = 128.] {
\label{fig:c}
\includegraphics[width=0.62\columnwidth]{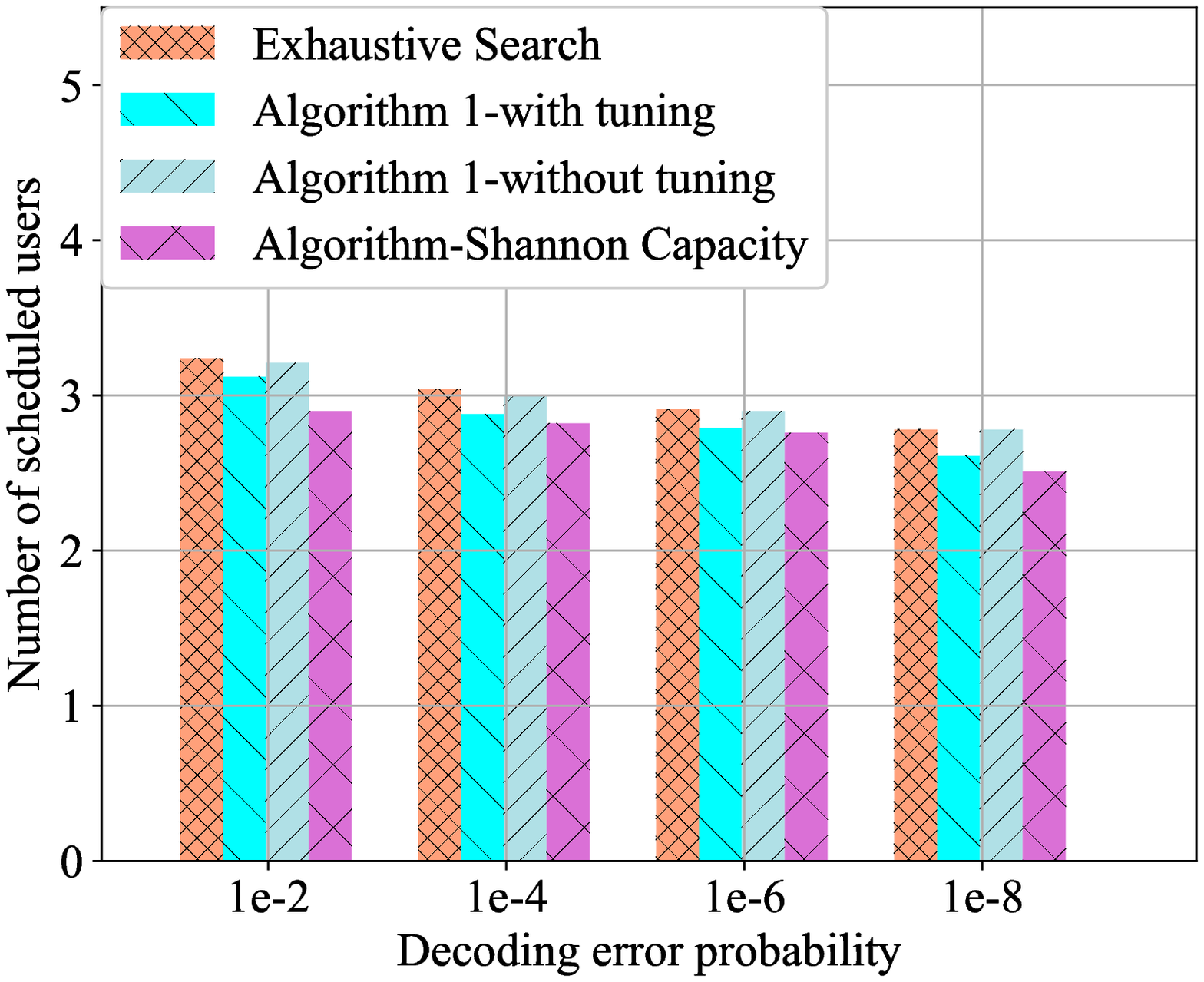}
}
\caption{Experiment results under different conditions. }
\label{fig1}
\end{figure*}

\begin{remark}
For convex problem~\eqref{URLLC13}, its optimal solution of the $\tau$-th iteration is also a feasible point of the problem in the $(\tau +1) $-th iteration. Therefore, Algorithm 1 is a non-increasing process. Besides, the power constraint~\eqref{URLLC06d} limits the lower bound of the problem and guarantees the convergence\cite{Tran2012}.
\end{remark}

\begin{remark}
Parameter $\mu$ penalizes the objective function for any $\kappa_{k}$ with its value between 0 and 1. A large value of $\mu$ will be helpful for convergence of the algorithm, but may miss some users. In this letter, a smaller value of $\mu$ would be chosen to find a better solution of \eqref{URLLC13} for the same initial solution.
\end{remark}
\begin{remark}
{An additional tuning process is adopted to solve the problem when $\kappa_{k} = 0, ||\mathbf{w}_{k}||^2 \ge 0$ even iteration stops, since the residual power is larger than zeros but insufficient for serving more users. The tuning process begins by choosing the $k$-th user with this situation. Then its initial beamforming vector is set to be 0, and the whole optimization steps 2-4 will be executed again. Here,  algorithm 1 with and without tuning operations are respectively referred as Algorithm 1 - with tuning and Algorithm 1 - without tuning.}
\end{remark}

\section{\label{simulation} Numerical Results}
In this section, we compare different algorithms on simulated data. Simulated channel follows the same model as in the reference \cite{TWCHe2020}, where channel coefficient ${{\bf{h}}_k} = \sqrt {1/{{\left( {1 + ({d_k}/{d_0}} \right)}^\varrho })} {{{\bf{\tilde h}}}_k}$. Here, elements of ${{{\bf{\tilde h}}}_k}$ are assumed to be independent and identically distributed, fading exponent $\varrho$ = 3, reference distance $d_{0}$ = $30$ m, cell radius is $300$ m. All users have the same noise variance, i.e., $\sigma_{k}^{2}=\sigma^{2}$, $\forall k\in\mathcal{K}$. SNR is defined as $ =10\log_{10}\left(\frac{P}{\sigma^{2}}\right)$ in dB for easy of notation. Minimum rate for each user is $\frac{D}{n}$.{Monte Carlo simulations are implemented and the final results are all the mean values.} Other {main} parameters are listed as follows:  $\sigma$ = $1$, {antenna number $N_{\mathrm{t}}$ = $4$}, data size $D$ = 256 bits and SNR = $10$ dB. We compare the proposed algorithm with exhaustive search (ES) method, with feasibility criterion as in~\cite[IV-D]{TWCHe2020}, and the method using conventional Shannon capacity. The latter problem could be realized via solving (13c) with replacing ${\tilde \gamma _k}$ with ${\tilde \gamma ^* _k} = {2^{\frac{D}{n}}} - 1$. Among them, computational complexity of ES method is $O(\sum\limits_{\hat k = 1}^K {C_K^{\hat k}{{(\hat k{N_t})}^3}(K - \hat k + 1)\hat k} )$, where ${C_K^{\hat k}}$ is combinatorial number. The others share the same lower computational complexity, namely, $O(({K}{N_t})^3 (4K+1))$.

{{The number of scheduled users, latency and reliability are three crucial factors in designing uRLLC system, which are respectively investigated in Fig. \ref{fig:a}, \ref{fig:b} and \ref{fig:c}. Among them, the reliability is described by decoding error probability $\epsilon$, while the latency could be reflected by blocklength $n$, since it is proportional to multiplication of blocklength $n$ and sampling interval $T_s$. In the figure, more users would be scheduled with the increase of number of candidate users, $n$ and $\epsilon$ for the four algorithms in general. Algorithm 1 with tuning obtains the  closest results compared with ES method, and its performance outperforms Algorithm 1 without tuning at the expense of computational complexity. It is also worth noting that some scheduled users using conventional Shannon capacity might be unavailable for transmitting, since the requirement of SINR for users are relaxed in such circumstance. In the experiments, those users are removed from scheduled user set before we plot figures. This also explains the phenomenon that the results using Shannon capacity are much worse than the proposed methods in Fig. \ref{fig:a} and Fig. \ref{fig:b}.}}

\section{\label{conclusion} Conclusions}
In this letter, maximizing the set cardinality of users scheduled is investigated for ultra-dense uRLLC networks. Instead of maximizing the system sum rate with Shannon rate, a joint US and BF optimization model, which aims at scheduling as many users as possible under the maximum allowable transmission power and the minimum rate requirement of each user, is proposed and effectively solved. Simulation results show the effectiveness of the proposed algorithm. {Note that the letter  provides a novel view for resource allocation in ultra-dense uRLLC networks, and the proposed algorithm could also be extended to serving all users by using the Round Robin scheduling or considering the user fairness problem. Nevertheless, how to schedule all users with taking the fairness of users under consideration beyond the discussion of this letter and it would be investigated in the future work.}

\begin{small}

\end{small}
\end{document}